\documentstyle[12pt,epsfig,graphicx]{article} 
\makeatletter \def\@cite#1#2{{[{#1}]\if@tempswa\typeout {IJCGA
warning: optional citation argument ignored: `#2'} \fi}}

\newcount\@tempcntc
\def\@citex[#1]#2{\if@filesw\immediate\write\@auxout{\string\citation{#2}}\fi
  \@tempcnta\z@\@tempcntb\m@ne\def\@citea{}\@cite{\@for\@citeb:=#2\do
    {\@ifundefined
       {b@\@citeb}{\@citeo\@tempcntb\m@ne\@citea\def\@citea{,}{\bf ?}\@warning
       {Citation `\@citeb' on page \thepage \space undefined}}%
    {\setbox\z@\hbox{\global\@tempcntc0\csname b@\@citeb\endcsname\relax}%
     \ifnum\@tempcntc=\z@ \@citeo\@tempcntb\m@ne
       \@citea\def\@citea{,}\hbox{\csname b@\@citeb\endcsname}%
     \else
     \advance\@tempcntb\@ne
      \ifnum\@tempcntb=\@tempcntc
      \else\advance\@tempcntb\m@ne\@citeo
      \@tempcnta\@tempcntc\@tempcntb\@tempcntc\fi\fi}}\@citeo}{#1}}
\def\@citeo{\ifnum\@tempcnta>\@tempcntb\else\@citea\def\@citea{,}%
  \ifnum\@tempcnta=\@tempcntb\the\@tempcnta\else
   {\advance\@tempcnta\@ne\ifnum\@tempcnta=\@tempcntb \else 
\def\@citea{--}\fi
    \advance\@tempcnta\m@ne\the\@tempcnta\@citea\the\@tempcntb}\fi\fi}
\makeatother

\def\boxit#1{\leavevmode\thinspace\hbox{\vrule\vtop{\vbox{\hrule%
        \vskip3pt\kern1pt\hbox{\vphantom{\bf/}\thinspace\thinspace%
        {\bf#1}\thinspace\thinspace}}\kern1pt\vskip3pt\hrule}\vrule}%
        \thinspace}
\def\Boxit#1{\noindent\vbox{\hrule\hbox{\vrule\kern3pt\vbox{
\advance\hsize-7pt\vskip-\parskip\kern3pt\bf#1 \hbox{\vrule height0pt
depth\dp\strutbox width0pt} \kern3pt}\kern3pt\vrule}\hrule}}

\newcommand{\gsim}{\lower.7ex\hbox{$\;\stackrel{\textstyle>}{\sim}\;$}}
\newcommand{\lsim}{\lower.7ex\hbox{$\;\stackrel{\textstyle<}{\sim}\;$}}
\newcommand{\be}{\begin{equation}} \newcommand{\ee}{\end{equation}}
\newcommand{\beq}{\begin{equation}} \newcommand{\eeq}{\end{equation}}
\newcommand{\bea}{\begin{eqnarray}} \newcommand{\eea}{\end{eqnarray}}

\newcommand{\mDM}{m_{\rm DM}}
\newcommand{\tDM}{\tau_{\rm DM}}

\newcommand{\Ed}{E_{\bar d}}

\def\baselinestretch{1}
\begin{document}
\catcode`@=11 \newtoks\@stequation
\def\subequations{\refstepcounter{equation}%
\edef\@savedequation{\the\c@equation}%
\@stequation=\expandafter{\theequation}
\edef\@savedtheequation{\the\@stequation}
\edef\oldtheequation{\theequation}
\def\theequation{\oldtheequation\alph{equation}}}
\def\endsubequations{\setcounter{equation}{\@savedequation}%
\@stequation=\expandafter{\@savedtheequation}%
\edef\theequation{\the\@stequation}\global\@ignoretrue

\noindent} \catcode`@=12
\begin{titlepage}

\title{\bf  Antideuterons from Dark Matter Decay}
\vskip3in \author{{\bf Alejandro Ibarra} and
{\bf David Tran\footnote{\baselineskip=16pt {\small E-mail addresses: {\tt
alejandro.ibarra@ph.tum.de, david.tran@ph.tum.de}}}}
\hspace{3cm}\\ \vspace{0.1cm}
{\normalsize\it  Physik-Department T30d, Technische Universit\"at M\"unchen,}\\[-0.05cm]
{\normalsize \it James-Franck-Stra\ss{}e, 85748 Garching, Germany.}
}  \date{}  \maketitle  \def\baselinestretch{1.15}
\begin{abstract}
\noindent
Recent observations of a large excess of cosmic-ray positrons at high 
energies have raised a lot of interest in leptonic decay 
modes of dark matter particles. Nevertheless, dark matter particles in 
the Milky Way halo could also decay hadronically,
producing not only a flux of antiprotons but also a flux of
antideuterons. We show that for certain choices of parameters
 the antideuteron flux from dark matter
decay can be much larger than the purely secondary flux from
spallation of cosmic rays on the interstellar medium, while 
the total antiproton flux remains consistent with present observations.
We show that if the dark matter particle is sufficiently light, 
the antideuteron flux from dark matter decay could even be within the reach of 
planned experiments such as AMS-02 or GAPS.
Furthermore, we discuss the prospects to observe the
antideuteron flux in the near future if the steep rise in the
positron fraction reported by the PAMELA collaboration is interpreted
in terms of the decay of dark matter particles.
\end{abstract}

\thispagestyle{empty}
\vspace*{0.2cm} \leftline{April 2009} \leftline{}

\vskip-17.0cm \rightline{TUM-HEP 718/09}

\vskip3in

\end{titlepage}
\setcounter{footnote}{0} \setcounter{page}{1}
\newpage
\baselineskip=20pt

\noindent

\section{Introduction}

There is mounting evidence for the existence of a hitherto undiscovered 
elementary particle
in the Universe, the dark matter particle, with an abundance which 
is approximately six times bigger than the abundance of baryons.
Dark matter particles are also known to exist in galactic halos
through their gravitational influence on the motion of 
stars~\cite{Bertone:2004pz}.  However, although dark matter particles 
are ubiquitous in the Universe at large, as well as in our own Galaxy, 
very little is known about their nature and properties.

One of the key questions about the dark matter properties 
is whether the dark matter particles are stable or not. 
Indeed, astrophysical and cosmological observations do not require 
the dark matter particles to be absolutely stable, but only to have
a lifetime larger than the age of the Universe. Furthermore, recent
evidence for the existence of a primary component in the high energy
electron and positron fluxes~\cite{Adriani:2008zr} could be naturally
accounted for by the decay of dark matter particles with a mass 
$m_{\rm DM}\gsim 300$ GeV, which decay preferentially into charged 
leptons of the first and  second generations with a lifetime 
$\tau_{\rm DM}\sim 10^{26}$s~\cite{Ibarra:2008jk}.

The properties of the dark matter particles are further constrained from 
observations of the diffuse gamma-ray flux by EGRET~\cite{Sreekumar:1997un}, 
as well as from observations of the antiproton flux by PAMELA~\cite{Adriani:2008zq}, BESS95~\cite{Matsunaga:1998he}, BESS95/97~\cite{Orito:1999re}, 
CAPRICE94~\cite{Boezio:1997ec}, CAPRICE98~\cite{Boezio:2001ac} and 
IMAX~\cite{Mitchell:1996bi}. More concretely,
the good agreement of the theoretical predictions for a purely secondary 
antiproton flux with the measurements indicates that the contribution
to the total antiproton flux from dark matter matter can only be subdominant.
On the other hand, the gamma-ray excess in the extragalactic background, 
revealed by the analysis in \cite{Strong:2004de} after 
subtracting the galactic foreground, hints at a possible exotic component
of diffuse gamma rays, which will be confirmed
or refuted by Fermi in the near future.

In this paper we will estimate the antideuteron flux from dark matter
decay as a complementary way to determine the properties of the
dark matter particle. As pointed out in~\cite{Donato:1999gy,Donato:2008yx}, 
for energies smaller than $\sim 3$ GeV the expectations for 
a purely secondary antideuteron flux, due to spallation 
of cosmic rays on the interstellar medium, lie below the present BESS 
limit on the  antideuteron flux~\cite{Fuke:2005it}, as well as the 
projected limits for AMS-02~\cite{Choutko,Koglin:2008zz} and 
GAPS~\cite{Hailey:2003xh,Hailey:2005yx}. Therefore, no detection of 
antideuterons is to be expected in these experiments if the antideuteron 
flux is of purely secondary origin.
At the same time, a discovery of cosmic antideuterons in the 
upcoming generation of
experiments would be strong evidence for a primary component 
and, if interpreted as the result of dark matter
decay, as evidence for hadronic decays of dark matter
particles, which may not be inferred from the antiproton flux due to
the large astrophysical background in this channel. 

In Section 2 we will review antideuteron production from dark
matter decay using the coalescence model. In Section 3 we will
review the propagation of antideuterons in the diffusive halo of the
Milky Way. In Section 4 we will show our results, and in Section
5 we will finally present our conclusions.

\section{Antideuteron Production}

We will assume that the Milky Way dark matter halo is 
populated by dark matter particles with mass $\mDM$, 
with their distribution following a 
density profile $\rho(\vec{r})$, where $\vec{r}$ denotes
position with respect to the center of the Galaxy.
The dark matter distribution is usually parametrized as a spherically
symmetric profile
\begin{equation}
\rho(r)=\frac{\rho_0}{(r/r_c)^\gamma 
[1+(r/r_c)^\alpha]^{(\beta-\gamma)/\alpha}}\;,
\end{equation}
where $r=|\vec{r}|$ and the parameters $\alpha$, $\beta$, $\gamma$ and
$r_c$ are listed in Table~\ref{tab:halomodels} for some commonly 
used halo models. Finally, $\rho_0$ is a parameter that is adjusted 
to yield a local halo density of
$\rho(r_\odot)=0.30\,{\rm GeV}/{\rm cm}^3$~\cite{Bergstrom:1997fj},
with $r_\odot = 8.5 ~\rm{kpc}$ being the distance of the Sun to the 
Galactic center.

\begin{table}[t]
\begin{center}
\begin{tabular}{|c|cccc|}
\hline
Halo model & $\alpha$ & $\beta$ & $\gamma$ & $r_c$ (kpc)\\
\hline
Navarro, Frenk, White~\cite{Navarro:1995iw}& 1 & 3 & 1 & 20\\

Isothermal & 2 & 2 & 0 & 3.5 \\
Moore~\cite{Moore:1999gc}& 1.5 & 3 & 1.5 &28\\
\hline
\end{tabular}
\caption{\label{tab:halomodels} \small 
Parameters characterizing some commonly used halo models.}
\end{center}
\end{table}

Dark matter particles at $\vec{r}$ eventually decay with lifetime $\tDM$,
producing antideuterons at a rate per unit energy and unit volume 
given by
\begin{equation}
Q_{\bar d}(\Ed,\vec{r})=
\frac{\rho(\vec{r})}{\mDM\tDM}\frac{dN_{\bar d}}{d\Ed}\;,
\label{source-term}
\end{equation}
where $dN_{\bar d}/d\Ed$ is the energy spectrum of antideuterons
produced in the decay.

The production of antideuterons in the fragmentation of weak gauge 
bosons or Higgs bosons can be described by the coalescence model
which assumes that the probability in momentum space of producing an 
antideuteron is proportional to the product of the probabilities of
producing a single antiproton and a single antineutron:
\begin{equation}
\left[\gamma_{\bar d} \frac{d^3N_{\bar d}}{d^3 \vec{k}_{\bar d}}\right]
=\int d^3 \vec{k}_{\bar p}d^3 \vec{k}_{\bar n} 
\;C(\vec{k}_{\bar p},\vec{k}_{\bar n})
\left[\gamma_{\bar p} \frac{d^3N_{\bar p}}{d^3 \vec{k}_{\bar p}}
\left(\vec{k}_{\bar p}\right)\right]
\left[\gamma_{\bar n} \frac{d^3N_{\bar n}}{d^3 \vec{k}_{\bar n}}
\left(\vec{k}_{\bar n}\right)\right]
\delta^{(3)}(\vec{k}_{\bar p}+\vec{k}_{\bar n}-\vec{k}_{\bar d})\;,
\end{equation}
where $\gamma$ is a Lorentz factor and $C(\vec{k}_{\bar p},\vec{k}_{\bar n})$ 
is the coalescence function which can only depend on the relative
momentum between antiproton and antineutron,
$\vec{\Delta}=\vec{k}_{\bar p}-\vec{k}_{\bar n}$. Using
that $\vec{k}_{\bar d}=\vec{k}_{\bar p}+\vec{k}_{\bar n}$, one
obtains
\begin{equation}
\left[\gamma_{\bar d} \frac{d^3N_{\bar d}}{d^3 \vec{k}_{\bar d}}\right]
=\int d^3 \vec{\Delta}
\;C(|\vec{\Delta}|)
\left[\gamma_{\bar p} \frac{d^3N_{\bar p}}{d^3 \vec{k}_{\bar p}}
\left( \vec{k}_{\bar p}=\frac{\vec{k}_{\bar d}+\vec{\Delta}}{2}\right)\right]
\left[\gamma_{\bar n} \frac{d^3N_{\bar n}}{d^3 \vec{k}_{\bar n}}
\left( \vec{k}_{\bar n}=\frac{\vec{k}_{\bar d}-\vec{\Delta}}{2}\right)\right]\;.
\end{equation}
The coalescence function is strongly peaked at 
$|\overrightarrow{\Delta}|\simeq 0$ since the binding energy
of the antideuteron, $B\simeq 2.2~{\rm MeV}$, is much smaller than 
its rest mass, $m_{\bar d}\simeq 1.88~{\rm GeV}$. 
Therefore, the previous
expression can be approximated by
\begin{equation}
\left[\gamma_{\bar d} \frac{d^3N_{\bar d}}{d^3 \vec{k}_{\bar d}}\right]
\simeq\int d^3 \vec{\Delta} \;C(|\vec{\Delta}|)
\left[\gamma_{\bar p} \frac{d^3N_{\bar p}}{d^3 \vec{k}_{\bar p}}
\left( \vec{k}_{\bar p}=\frac{\vec{k}_{\bar d}}{2}\right)\right]
\left[\gamma_{\bar n} \frac{d^3N_{\bar n}}{d^3 \vec{k}_{\bar n}}
\left( \vec{k}_{\bar n}=\frac{\vec{k}_{\bar d}}{2}\right)\right]\;.
\label{eq:derivation3}
\end{equation}

It is useful to define the coalescence momentum, $p_0$, as
\begin{equation}
\int d^3 \vec{\Delta} \;C(|\vec{\Delta}|)\equiv\frac{4 \pi}{3} p^3_0\;,
\end{equation}
which can be interpreted as the maximum relative 
momentum between the antiproton and the antineutron 
for which an antideuteron will form by fusion of the two nucleons. 
The coalescence momentum can be determined experimentally
from proton--nucleus collisions, yielding 
$p_0\simeq 79~{\rm MeV}$\footnote{The 
uncertainty in the measurement of the hadronic production cross 
sections translates into a range for the coalescence momentum
that was estimated to be 
$p_0=79^{+26}_{-13}~{\rm MeV}$~\cite{Donato:2008yx}.}~\cite{Duperray:2005si}, 
or from 
$e^+ e^-$ collisions at the $Z^0$ resonance,
yielding $p_0\simeq 71.8\pm 3.6~{\rm MeV}$~\cite{Schael:2006fd}.
Since the decay of weak gauge bosons is precisely
the source of antideuterons in the decaying dark matter
scenario, we will adopt the latter value $p_0\simeq 71.8\pm 3.6~{\rm MeV}$
in the remainder of the paper. Note that the measured coalescence momentum
is not far from the estimate obtained by using the antideuteron binding energy, 
namely $\sqrt{m_{\bar p} B}\sim 46$ MeV, supporting the above physical
interpretation of the coalescence momentum.

The dark matter decay produces antideuterons,
antiprotons and antineutrons with an isotropic distribution,
and consequently
\begin{equation}
\left[\gamma \frac{d^3N}{d^3 \vec{k}}\right]
=\frac{1}{4\pi m k} \frac{dN}{dE}\;.
\end{equation}
Therefore, Eq.~(\ref{eq:derivation3}) can be cast as:
\begin{equation}
\frac{dN_{\bar d}}{d\Ed}\simeq
\frac{4 p_0^3}{3 k_{\bar d}} \frac{m_{\bar d}}{m_{\bar p} m_{\bar n}}
\left[\frac{dN_{\bar p}}{d E_{\bar p}}
\left(E_{\bar p}=\frac{\Ed}{2}\right)\right]^2\;,
\end{equation}
where we have assumed, employing isospin invariance, that the 
probability of producing
an antiproton with momentum $\vec{k_{\bar p}}$ in the fragmentation 
is the same as the probability of producing an antineutron.

\section{Propagation}

Antideuteron propagation in the Milky Way can be described by
a stationary two-zone diffusion model with cylindrical boundary 
conditions~\cite{ACR}. Under this approximation, 
the number density of antideuterons
per unit energy, $f_{\bar d}(E,\vec{r},t)$, 
satisfies the following transport equation:
\begin{equation}
0~=~\frac{\partial f_{\bar d}}{\partial t}~=~\vec{\nabla} 
\cdot [K(E,\vec{r}){\vec \nabla} f_{\bar d} - 
\vec{V}_c(\vec{r})  f_{\bar d}]
+Q(E,\vec{r})\;.
\label{transport}
\end{equation}
where we have neglected non-annihilating
interactions of antideuterons with the interstellar gas,
as well as the changes in the energy of the antideuteron
during their propagation due to energy losses
and reacceleration.\footnote{This assumption is well justified 
for antideuterons from dark matter decay, since after being produced 
in the Milky Way halo, they rarely cross the disk before reaching the
Earth.} The boundary conditions require the solution 
$f_{\bar d}(E,\vec{r},t)$ to vanish at the boundary
of the diffusion zone, which is approximated by a cylinder with 
half-height $L = 1-15~\rm{kpc}$ and radius $ R = 20 ~\rm{kpc}$.

The first term on the right-hand side of the transport equation
is the diffusion
term, which accounts for the propagation through the
tangled Galactic magnetic field.
The diffusion coefficient $K(E,\vec{r})$ is assumed to be constant
throughout the diffusion zone and is parametrized by:
\begin{equation}
K(E)=K_0 \;\beta\; {\cal R}^\delta\;,
\end{equation}
where
$\beta=v/c$ and ${\cal R}$ is the antideuteron rigidity, which
is defined as the momentum in GeV per unit charge, 
${\cal R}\equiv p({\rm GeV})/Z$.
The normalization $K_0$ and the spectral index $\delta$
of the diffusion coefficient are related to the properties 
of the interstellar medium and can be determined from the 
flux measurements of other cosmic ray species, mainly from 
the Boron-to-Carbon (B/C) ratio~\cite{Maurin:2001sj}. 

The second term is the convection term, which accounts for
the drift of charged particles away from the 
disk induced by the Milky Way's Galactic wind. 
It has axial direction and is also assumed to be constant
inside the diffusion region: 
$\vec{V}_c(\vec{r})=V_c\; {\rm sign}(z)\, \vec{e}_z$.
The third term accounts for antimatter annihilation 
when it interacts with ordinary matter in the Galactic disk,
which is assumed to be an ``infinitely'' thin disk with half-width
$h=100$ pc. The annihilation rate, $\Gamma_{\rm ann}$, is given by:
\begin{equation}
\Gamma_{\rm ann}=(n_{\rm H}+4^{2/3} n_{\rm He})
\sigma^{\rm ann}_{\bar d p} v \;.
\end{equation}
In this expression it has been assumed that the annihilation cross
section between an antideuteron and a helium nucleus is
related to the annihilation cross section between an
antideuteron and a proton by the simple geometrical factor $4^{2/3}$.
On the other hand, $n_{\rm H}\sim 1\;{\rm cm}^{-3}$ is the number
density of Hydrogen nuclei in the Milky Way disk,
$n_{\rm He}\sim 0.07 ~n_{\rm H}$ the number density
of Helium nuclei and $\sigma_{\rm ann}^{\bar d p}$ is
the annihilation cross section. No experimental data are available
for the ${\bar d} p$ collisions, although
measurements exist for the total cross
section of the charge-conjugated process $d {\bar p} \rightarrow X$.
Since strong interactions preserve charge conjugation, it is 
reasonable to assume that $\sigma^{\rm tot}_{{\bar d} p}=
\sigma^{\rm tot}_{d {\bar p}}$. However, it is not the
total cross section that is required to compute the depletion
of antideuterons during their propagation, but the annihilation
cross section. Unfortunately, there is no experimental information
about the annihilation cross section neither for ${\bar d} p$ nor
for $d {\bar p}$ collisions. However, it was noted in~\cite{Donato:2008yx}
that the total cross section for $d {\bar p} \rightarrow X$ 
can be well approximated, within $\sim$ 10\%, by 
$2\sigma^{\rm tot}_{p{\bar p}}$ and  hence 
$\sigma^{\rm tot}_{{\bar d} p}\simeq 2\sigma^{\rm tot}_{p{\bar p}}$.
Then, assuming that the same rule applies for the annihilation
cross section, one obtains  $\sigma^{\rm ann}_{{\bar d} p}\simeq 
2\sigma^{\rm ann}_{p{\bar p}}$. For our numerical analysis,
we will adopt the parametrization by Tan and Ng of the 
proton-antiproton annihilation cross section~\cite{Tan:1983de}:
\begin{eqnarray}
\sigma^{\rm ann}_{p \bar p }(T_{\bar p}) = \left\{
\begin{array}{ll}
661\;(1+0.0115\;T_{\bar p}^{-0.774}-0.948\;T_{\bar p}^{0.0151})
\; {\rm mbarn}\;,
 & T_{\bar p} < 15.5\;{\rm GeV}~, \\
36 \;T_{\bar p}^{-0.5}\; {\rm mbarn}\;, & T_{\bar p} \geq 15.5\;{\rm GeV}\,, \\
\end{array} \right. 
\end{eqnarray}
where $T_{\bar p}$ is the kinetic energy of the antiproton.

The solution of the transport equation at the Solar System, 
$r=r_\odot$, $z=0$, can be formally expressed by the convolution
\begin{equation}
f_{\bar d}(\Ed)=\frac{1}{m_{\rm DM} \tau_{\rm DM}}
\int_0^{m_{\rm DM}}d\Ed^\prime G_{\bar d}(\Ed,\Ed^\prime) 
 \frac{dN_{\bar d}(\Ed^\prime)}{d\Ed^\prime}\;.
\label{solution}
\end{equation}
The analytic expression for the 
Green's function reads~\cite{Donato:2001sr}:
\begin{equation}
G_{\bar d}(T,T^\prime)=\sum_{i=1}^{\infty}
{\rm exp}\left(-\frac{V_c L}{2 K(T)}\right)
\frac{y_i(T)}{A_i(T) {\rm sinh}(S_i(T) L/2)} 
J_0\left(\zeta_i \frac{r_{\odot}}{R}\right)\delta(T-T^\prime)\;,
\end{equation}
where
\begin{equation}
y_i(T)=\frac{4}{J^2_1(\zeta_i)R^2}
\int_0^R r^\prime \,dr^\prime\; J_0\left(\zeta_i \frac{r^\prime}{R}\right) 
\int_0^L dz^\prime {\rm exp}
\left(\frac{V_c (L-z^\prime)}{2 K(T)}\right)
{\rm sinh}\left(\frac{S_i(L-z^\prime)}{2}\right)
\rho(\vec{r}\,^\prime)\;,
\end{equation}
and
\begin{eqnarray}
A_i(T)&=&2 h \Gamma_{\rm ann}(T) + V_c+k S_i(T) 
{\rm coth} \frac{S_i(T) L}{2}\;,\\
S_i(T)&=&\sqrt{\frac{V_c^2}{K(T)^2}+\frac{4 \zeta_i^2}{R^2}}\;.
\end{eqnarray}
We find that the Green's function can be numerically 
approximated by the following interpolation function:
\begin{equation}
G_{\bar d}(T,T^\prime)\simeq 10^{14}\,
e^{x +y \ln T +z \ln^2T}
\delta(T^\prime-T)\,{\rm cm}^{-3}\,{\rm s}\;,
\label{interp-antip}
\end{equation}
which is valid for any decaying dark matter particle. The coefficients
$x$, $y$ and $z$ for the NFW profile can be found in 
Table~\ref{tab:fit-antideuteron}
for the various diffusion models in Table~\ref{tab:param-antideuteron};
the dependence of the Green's function on the halo model is fairly weak.
In this case the approximation is better than about 5\% for
$T_{\bar d}$ between 1 and 100 GeV/n and better than about 20\% for 
$T_{\bar d}$ between 0.1 and 1 GeV/n.

\begin{table}[t]
\begin{center}
\begin{tabular}{|c|cccc|}
\hline
Model & $\delta$ & $K_0\,({\rm kpc}^2/{\rm Myr})$ & $L\,({\rm kpc})$
& $V_c\,({\rm km}/{\rm s})$ \\
\hline 
MIN & 0.85 & 0.0016 & 1 & 13.5 \\
MED & 0.70 & 0.0112 & 4 & 12 \\
MAX & 0.46 & 0.0765 & 15 & 5 \\
\hline
\end{tabular}
\caption{\label{tab:param-antideuteron} \small 
Astrophysical parameters compatible with the B/C ratio that 
yield the minimal (MIN), median (MED) and maximal (MAX) flux of antideuterons.}
\end{center}
\end{table}
\begin{table}[t]
\begin{center}
\begin{tabular}{|c|ccc|}
 \hline
model & $x$ & $y$ & $z$ \\ 
\hline
MIN &  $-$0.3889 &	0.7532	   &	$-$0.1788 \\
MED &	1.6023	 &	0.4382	   &	$-$0.1270 \\
MAX &	3.1992   &	$-$0.1098  &	$-$0.0374 \\
\hline
\end{tabular}
\caption{\label{tab:fit-antideuteron}\small 
Coefficients of the interpolating function Eq.~(\ref{interp-antip}) 
for the antideuteron Green's function for the NFW halo profile.}
\end{center}
\end{table}

Finally, the flux of primary antideuterons at the Solar System
from dark matter decay is given by:
\begin{equation}
\Phi_{\bar d}^{\rm{prim}}(\Ed) = \frac{v}{4 \pi} f_{\bar d}(\Ed),
\label{flux}
\end{equation}
where $v$ is the velocity of the antideuteron. 

However, this is not the antideuteron flux measured by balloon
or satellite experiments, which is affected by solar modulation.
In the force field approximation~\cite{solar-modulation} 
the effect of solar modulation can be included
by applying the following simple formula that relates 
the antideuteron flux at the top of the Earth's atmosphere and
the interstellar antideuteron flux~\cite{perko}:
\begin{equation}
\Phi_{\bar d}^{\rm TOA}(T_{\rm TOA})=
\left(
\frac{2 m_{\bar d} T_{\rm TOA}+T_{\rm TOA}^2}
{2 m_{\bar d} T_{\rm IS}+T_{\rm IS}^2}
\right)
\Phi_{\bar d}^{\rm IS}(T_{\rm IS}),
\end{equation}
where $T_{\rm IS}=T_{\rm TOA}+\phi_F$, with
$T_{\rm IS}$ and $T_{\rm TOA}$ being the antideuteron kinetic energies 
at the heliospheric boundary and at the top of the Earth's atmosphere,
respectively, and $\phi_F$ being the solar modulation parameter,
which varies between 500 MV and 1.3 GV over the eleven-year solar
cycle. Since experiments are usually undertaken near
solar minimum activity, we will choose $\phi_F=500$ MV 
for our numerical analysis in order to compare our predicted flux with 
the collected data. 

\section{Antideuteron Flux at Earth from Dark Matter Decay}

Using the injection spectrum of antideuterons from dark matter decay
calculated in Section 2 and the propagation formalism described in 
Section 3, it is straightforward to calculate the antideuteron flux
at Earth from dark matter decay.\footnote{For calculations of
the antideuteron flux from the annihilation of dark matter particles,
see \cite{Donato:1999gy,Donato:2008yx,Braeuninger:2009pe}.}
We will pursue here a model-independent
approach, calculating the expected antideuteron flux at Earth for 
various hadronic decay channels and different dark matter masses. To better
compare the predictions for the different possibilities, we will
first fix the dark matter lifetime to be $\tau_{\rm DM}=10^{26}$s,
which is the order of magnitude which could explain the HEAT and
PAMELA anomalies in the positron fraction \cite{Ibarra:2008qg,Ishiwata:2008cu,Ibarra:2008jk,positron-PAMELA} and which saturates the EGRET constraints on the 
diffuse gamma-ray flux~\cite{gammas,Ishiwata:2008cu}. Later on, we will 
discuss in more detail the predictions for the antideuteron flux in the 
light of the PAMELA anomaly. 

In the case that the dark matter particle is a fermion $\psi$,
the following hadronic decay channels are possible:
\begin{eqnarray}
\psi&\rightarrow& Z^0 \nu\;, \nonumber \\ 
\psi&\rightarrow& W^\pm \ell^\mp\;, \nonumber\\
\psi&\rightarrow& h^0 \nu\;.
\end{eqnarray}
The fragmentation of the weak gauge bosons and the Standard Model Higgs
boson produces a flux of antideuterons which has been calculated using 
the coalescence model to simulate the nuclear fusion of an 
antiproton and an antineutron ({\it cf.} Section 2), and the
event generator PYTHIA 6.4~\cite{Sjostrand:2006za}, 
to simulate the production of individual antiprotons and antineutrons.

Clearly, an exotic contribution to the antideuteron flux is correlated 
with an exotic contribution to the antiproton flux, which is severely
constrained by a number of experiments. Namely, the measurements of the
antiproton flux by  PAMELA~\cite{Adriani:2008zq}, 
BESS95~\cite{Matsunaga:1998he}, BESS95/97~\cite{Orito:1999re}, 
CAPRICE94~\cite{Boezio:1997ec}, CAPRICE98~\cite{Boezio:2001ac} and 
IMAX~\cite{Mitchell:1996bi} do not show any significant 
deviation from the predictions by conventional astrophysical 
models of spallation of cosmic rays on the Milky Way disk.
Therefore, in order to evaluate the prospects of observing the
antideuteron flux in future experiments, one has to ensure that
the prediction for the total antiproton flux does not exceed the
observed flux. The antiproton flux from dark matter decay has
been calculated following the analysis in \cite{Ibarra:2008qg},
for the same propagation models as for the antideuteron flux.

\begin{figure}[t]
\begin{center}
\begin{tabular}{c}
\psfig{figure=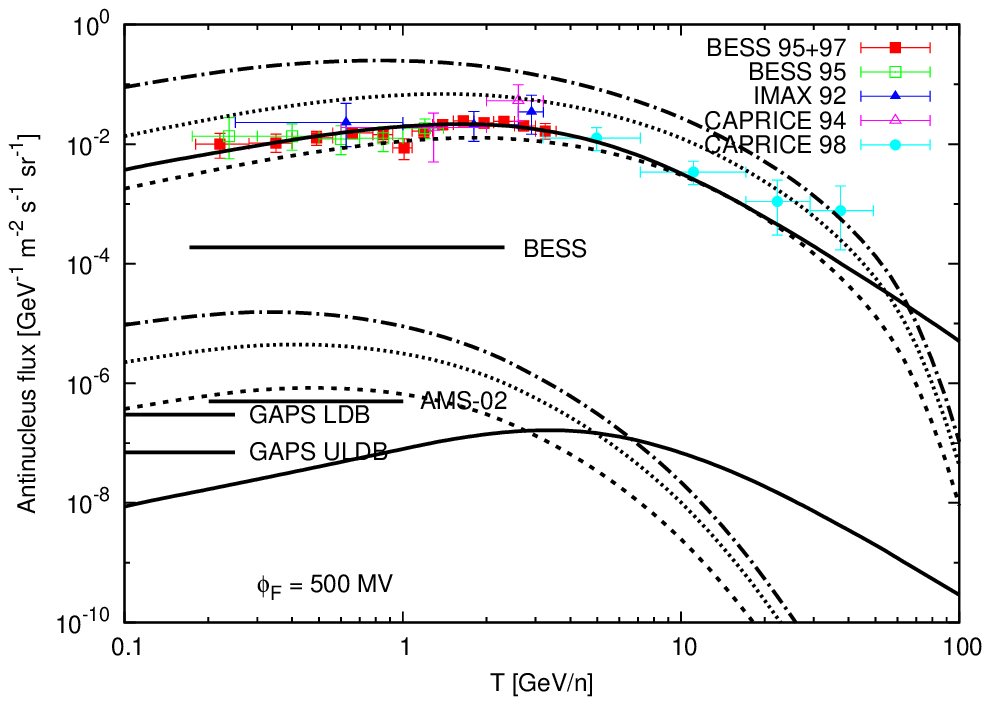,width=75mm} 
\psfig{figure=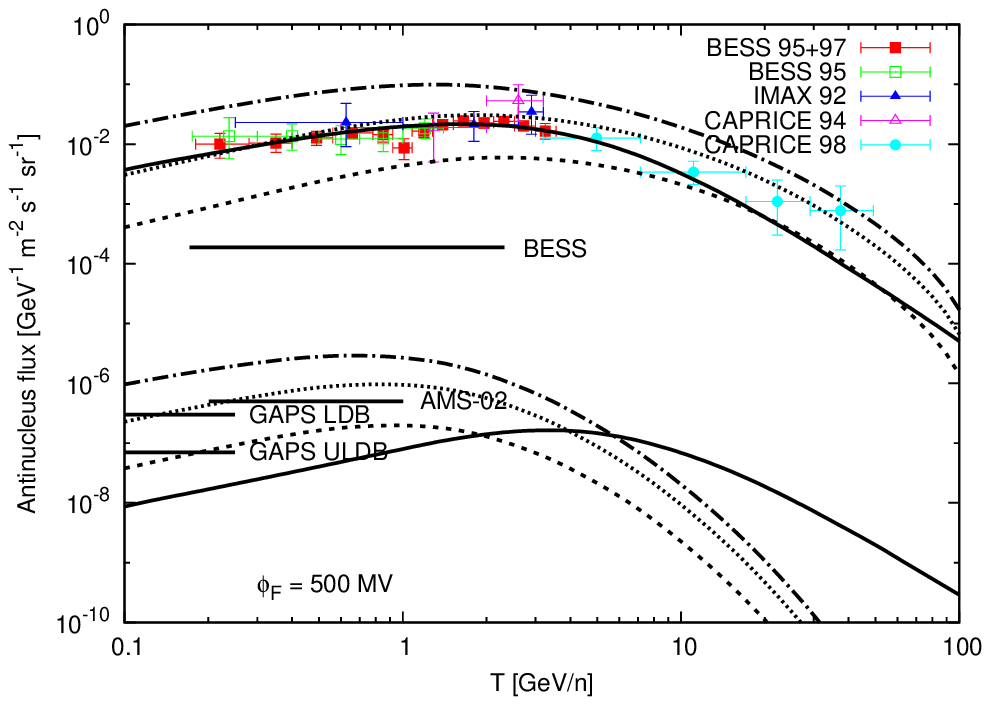,width=75mm} \\
\psfig{figure=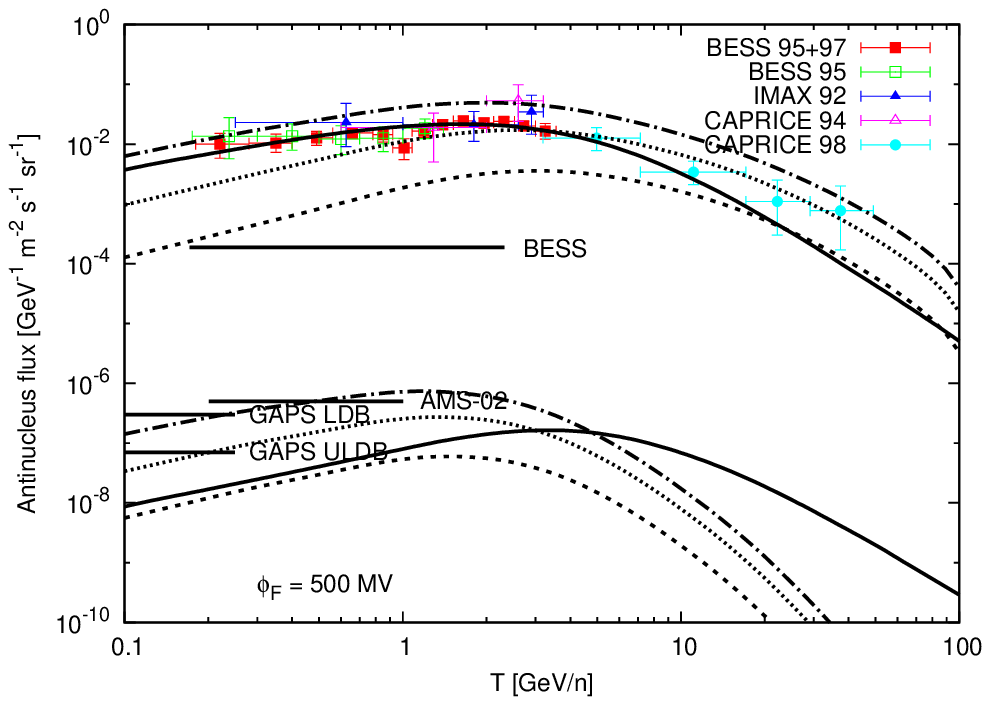,width=75mm} 
\psfig{figure=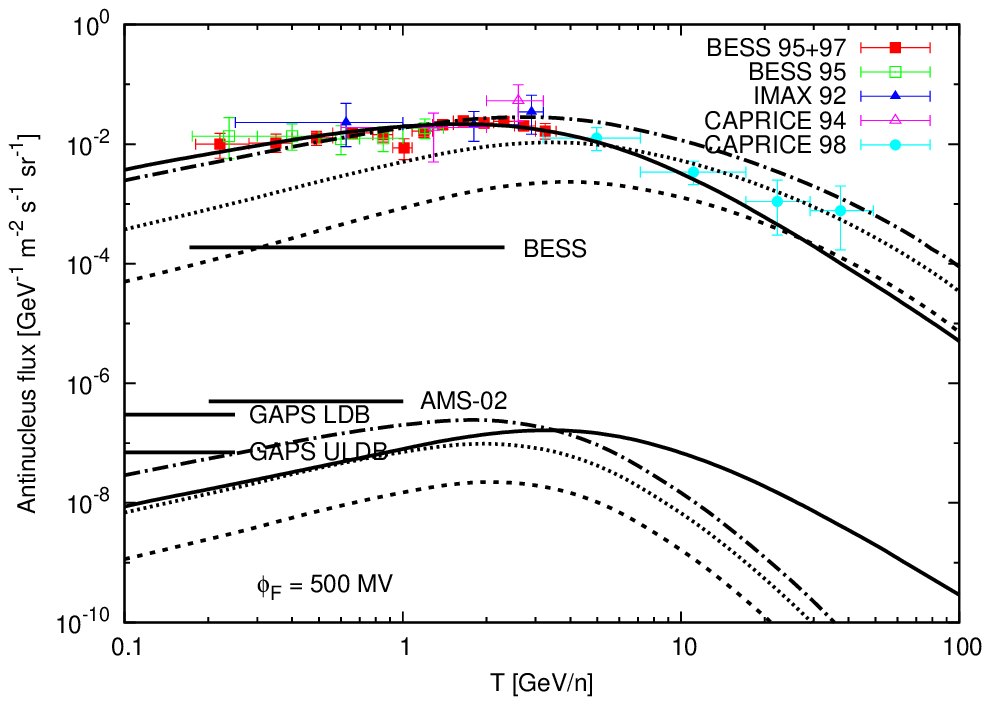,width=75mm} \\
\end{tabular}
\end{center}
\caption{\label{Znu}\small
Fluxes of antiprotons (upper curves) and antideuterons (lower curves)
from the decay of fermionic dark matter particles, assuming that the 
dark matter particle decays exclusively as $\psi \rightarrow Z^0 \nu$ 
with a fixed lifetime $\tau_{\rm DM} = 10^{26}$s. The dark matter mass is
200 GeV (top left panel), 400 GeV (top right), 600 GeV (bottom left)
and 800 GeV (bottom right). The dashed lines indicate the primary fluxes
from dark matter decay for the MIN propagation model, the dotted lines 
for the MED propagation model, 
and the dash-dotted lines for the MAX propagation model 
({\it cf.} Table \ref{tab:param-antideuteron}). The solid lines 
indicate the secondary fluxes.}
\end{figure}

\begin{figure}[t]
\begin{center}
\begin{tabular}{c}
\psfig{figure=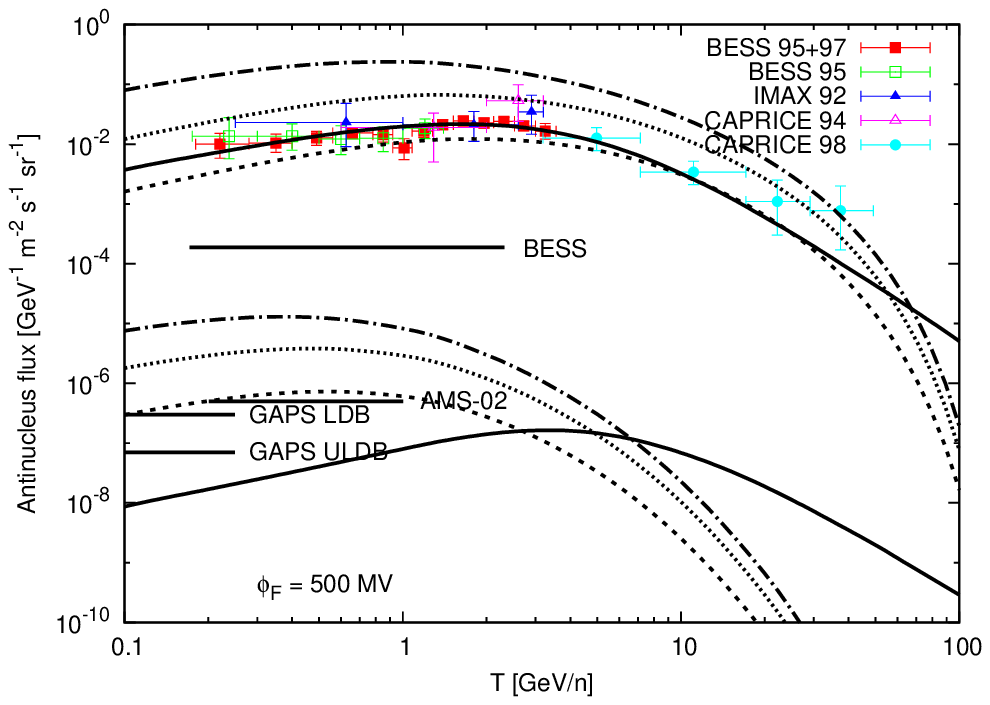,width=75mm} 
\psfig{figure=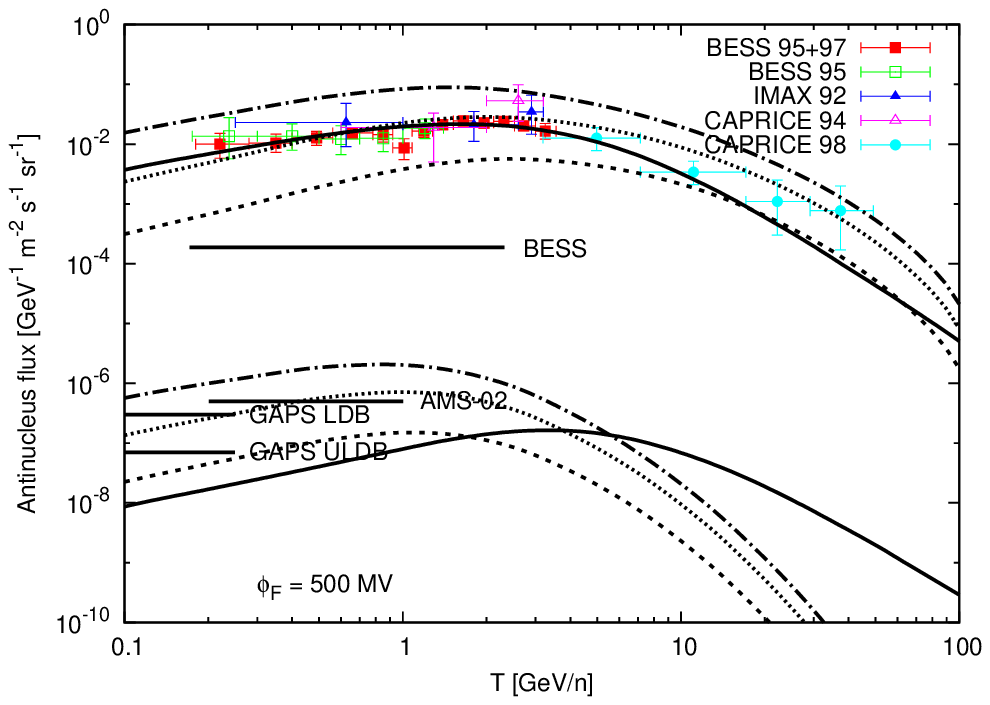,width=75mm} \\
\psfig{figure=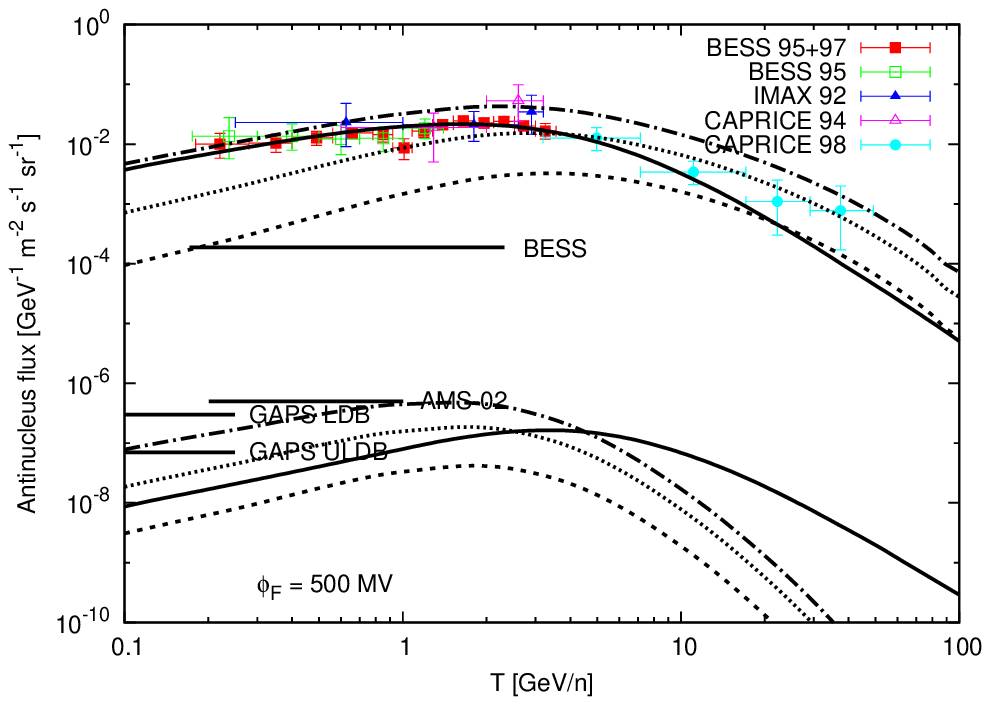,width=75mm} 
\psfig{figure=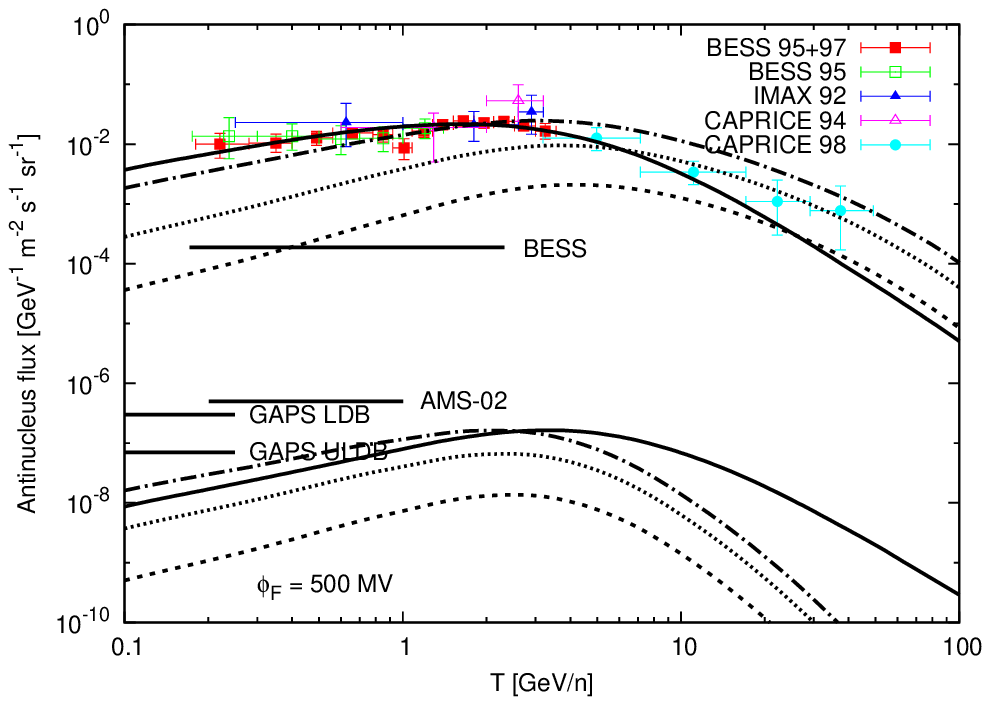,width=75mm} \\
\end{tabular}
\end{center}
\caption{\label{Wlep}\small 
The same as Fig.~\ref{Znu} for the decay mode  
$\psi \rightarrow W^\pm \ell^\mp$.}
\end{figure}

\begin{figure}[t]
\begin{center}
\begin{tabular}{c}
\psfig{figure=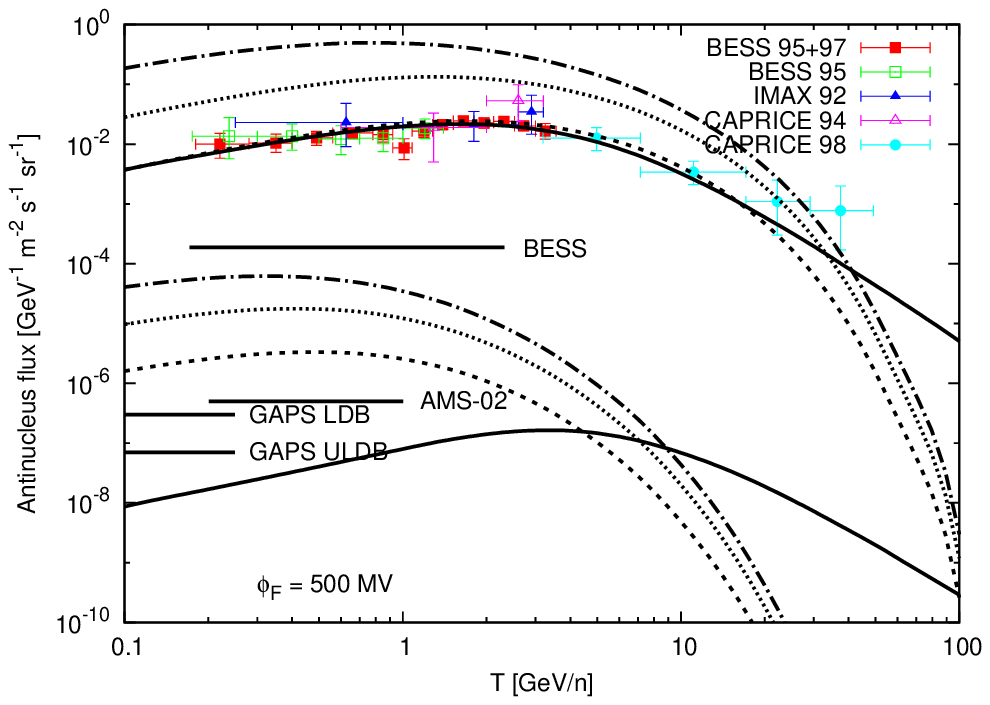,width=75mm} 
\psfig{figure=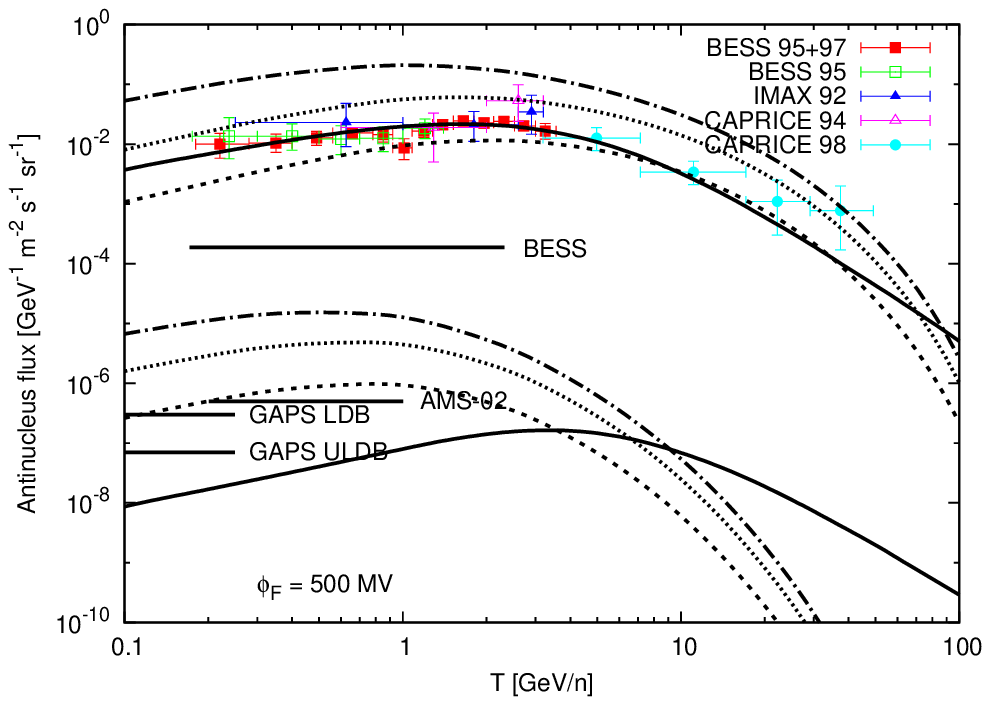,width=75mm} \\
\psfig{figure=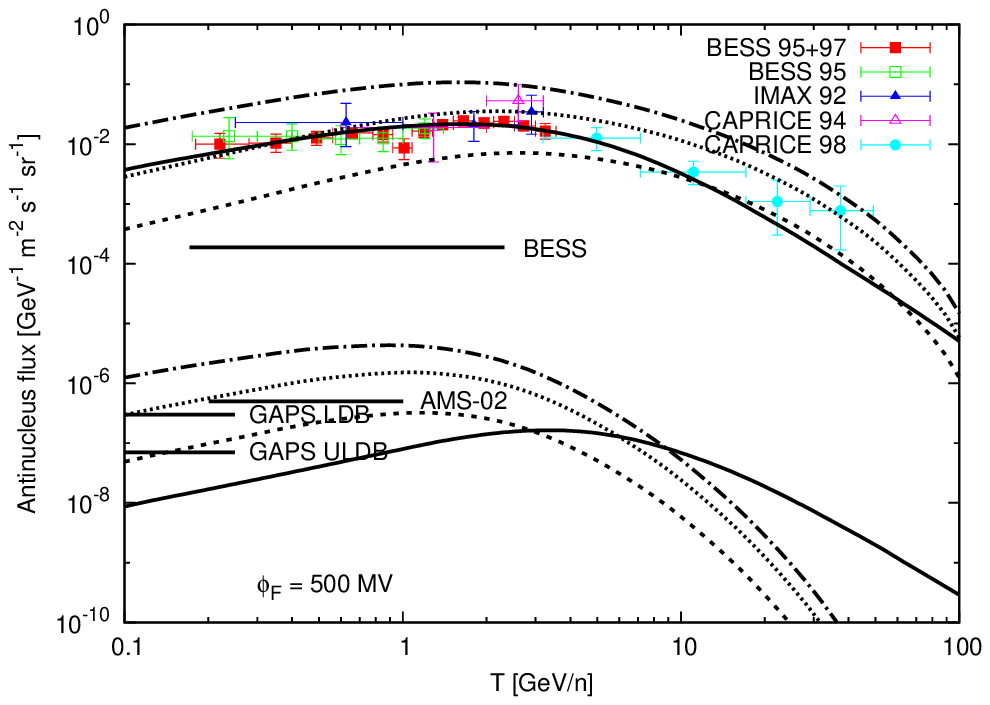,width=75mm} 
\psfig{figure=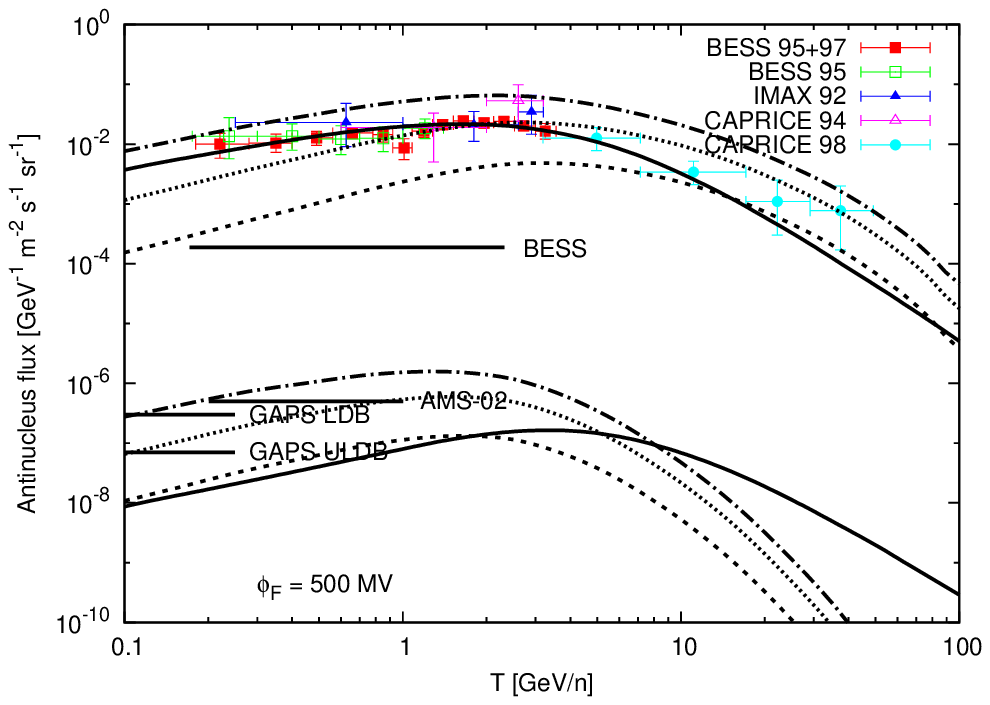,width=75mm} \\
\end{tabular}
\end{center}
\caption{\label{hnu}\small 
The same as Fig.~\ref{Znu} for the decay mode  
$\psi \rightarrow h^0 \nu$.}
\end{figure}

We show in Figs.~\ref{Znu},~\ref{Wlep},~\ref{hnu} the antiproton
and antideuteron fluxes from dark matter decay as  a function
of the kinetic energy per nucleon, assuming that
the dark matter particle is a fermion $\psi$, which decays exclusively
into $Z^0 \nu$, $W^\pm \ell^\mp$ or $h^0 \nu$, respectively. The lifetime
has been fixed to $\tau_{\rm DM}=10^{26}$s, while the dark matter mass has been
chosen to be $m_{\rm DM}=200,~ 400,~ 600,~ 800$ GeV. 
The predictions for the antiproton
and antideuteron fluxes for a general model with a different dark matter
lifetime and with arbitrary branching ratios can be straightforwardly
derived from these figures. In each plot, we show the prediction for the antiproton and antideuteron fluxes at the top of the atmosphere for the 
MIN, MED and MAX propagation models (see table \ref{tab:param-antideuteron}) 
as well as the expected flux from spallation, taken 
from \cite{Donato:2001sr} in the case of the antiprotons
and from \cite{Donato:2008yx} in the case of the antideuterons.\footnote{In 
these plots we only show the prediction for the secondary antiproton 
and antideuteron fluxes for the MED propagation model.
Whereas the uncertainty on the propagation model is typically not 
very important for the prediction of the secondary fluxes, there
exists a more important source of uncertainty
stemming from  the nuclear
and hadronic cross sections, which can be as large as 25\% for
antiprotons~\cite{Donato:2001sr} and 100\% for 
antideuterons~\cite{Donato:2008yx}.} For the halo model, we adopted 
the Navarro-Frenk-White (NFW) profile; the results for other halo profiles
are very similar to the ones presented here. 
We also show the present upper limit on the antideuteron flux from 
BESS~\cite{Fuke:2005it}, as well as the projected limits 
for AMS-02 for three years of data taking~\cite{Choutko,Koglin:2008zz}, as 
well as for 
GAPS~\cite{Hailey:2003xh,Hailey:2005yx} for a long duration
balloon (LDB) flight (60 days total over three flights) 
and for an ultra-long duration balloon (ULDB) flight (300 days total).

For this particular choice of the lifetime, $\tau_{\rm DM}=10^{26}$ s, 
we find wide ranges of diffusion parameters yielding a total
antiproton flux consistent with the observations. Besides, the total
antideuteron flux always lies well below the present BESS bound. We also 
find that for dark matter masses below $\sim 1$ TeV, the antideuteron flux
from dark matter decay can be, at energies below 3 GeV, 
significantly larger than the secondary antideuteron flux from spallation. 
Furthermore, the primary antideuteron flux from dark matter decay could be
large enough to be observable at the projected experiments AMS-02 and GAPS.
Therefore, since the purely secondary antideuteron flux is expected to
be below the sensitivity of projected experiments, the observation of 
an antideuteron flux in the near future could be interpreted as 
a signature of dark matter particles which decay hadronically.

When the decaying dark matter particle is a scalar,
the following hadronic decay modes are possible:
\begin{eqnarray}
\phi&\rightarrow& Z^0 Z^0, \nonumber \\ 
\phi&\rightarrow& W^+ W^-,  \nonumber\\
\phi&\rightarrow& h^0 h^0.
\end{eqnarray}
In these cases, the produced antideuteron flux is approximately 
twice as large as in the fermionic channels, $\psi\rightarrow Z^0 \nu$,
$\psi\rightarrow W^\pm \ell^\mp$,  $\psi\rightarrow h^0 \nu$,
respectively, and will not be discussed further. 

Let us now discuss the prospects for the detection of an antideuteron
flux in projected experiments, assuming that dark matter decay is 
the explanation of the positron excess reported by the PAMELA and HEAT
collaborations. As discussed in \cite{Ibarra:2008jk}, the steep rise
of the positron fraction measured by PAMELA can be explained by the
decay of a dark matter particle into hard electrons and positrons.
Namely, in the case that the dark matter particle is a fermion, 
the decay modes $\psi \rightarrow e^+ e^- \nu$ and
$\psi \rightarrow W^\pm e^\mp$ (and the analogous decay modes into muon
flavor) are favored by the data.  On the other hand, when the dark
matter particle is a scalar, the decay modes $\phi\rightarrow e^+ e^-,
~\mu^+ \mu^-$ are favored. Purely leptonic decay modes,  such as 
$\psi \rightarrow \ell^+ \ell^- \nu$ or $\phi\rightarrow \ell^+ \ell^-$
do not produce antideuterons. However, as discussed above, the decay
of a fermionic dark matter particle into $W^\pm$ bosons and charged leptons
could produce an observable antideuteron flux.

\begin{figure}[p]
\begin{center}
\begin{tabular}{c}
\psfig{figure=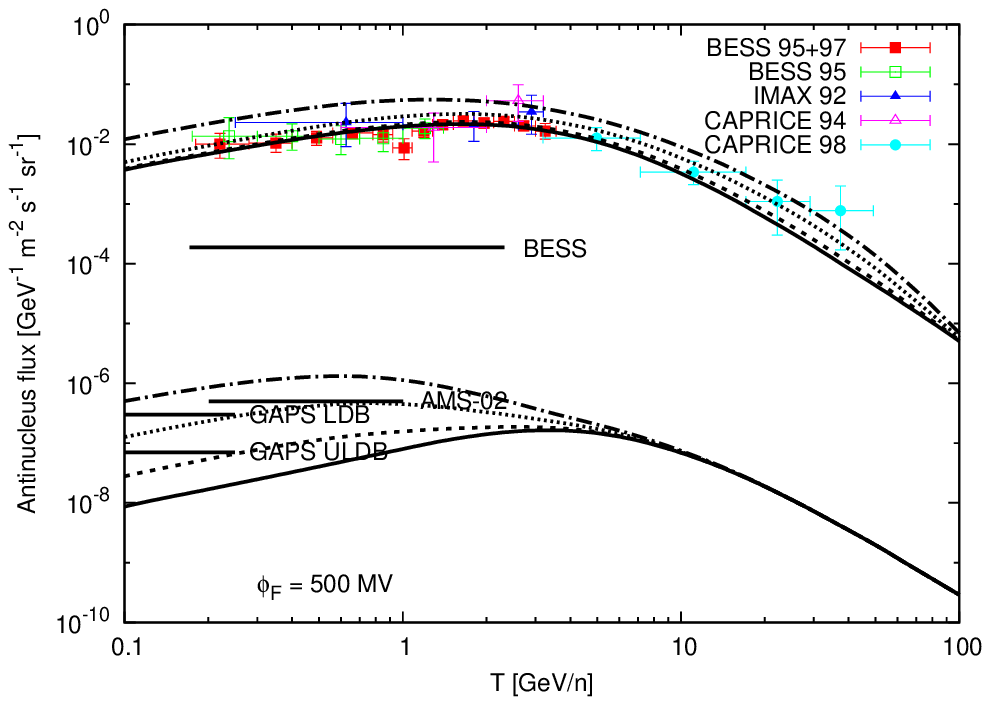,width=75mm}
\psfig{figure=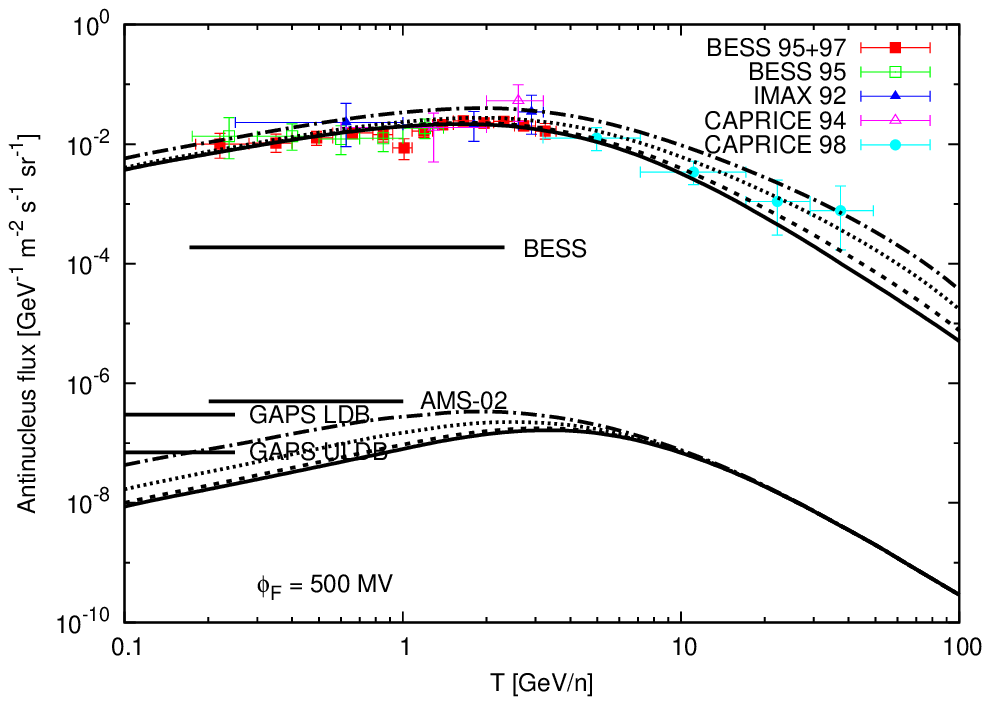,width=75mm} \\
\psfig{figure=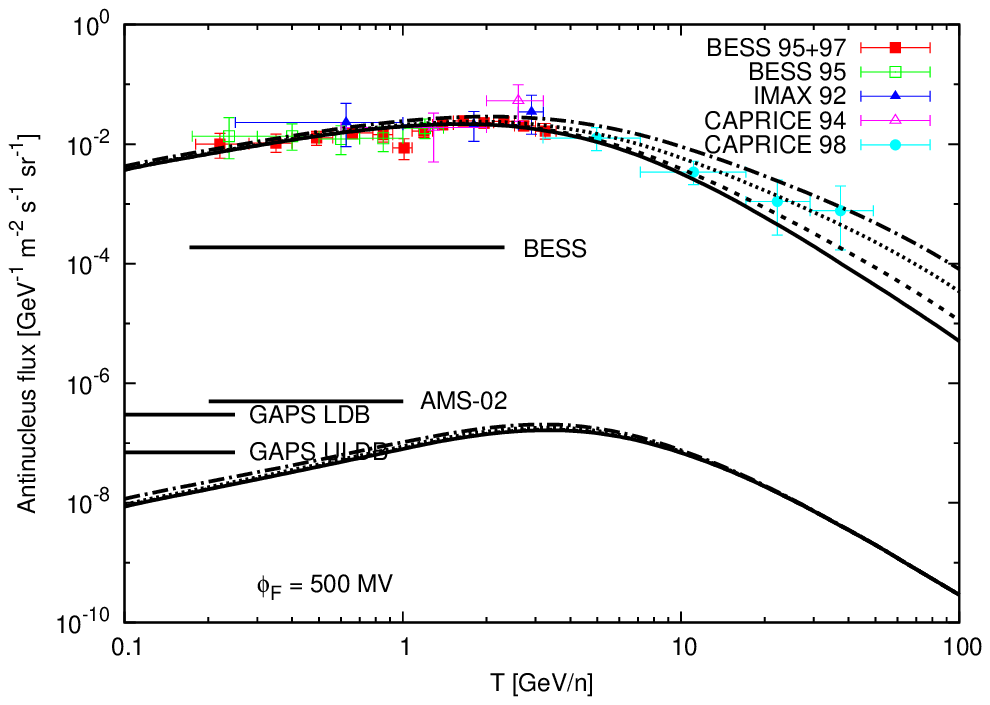,width=75mm} 
\end{tabular}
\end{center}
\caption{\label{WePAMELA}\small 
Total antiproton and antideuteron fluxes including a primary contribution
to the flux from the decay of a fermionic dark matter particle 
in the channel
$\psi\rightarrow W^\pm e^\mp$. The lifetime has been chosen, for each 
dark matter mass, to reproduce the steep rise in the positron
fraction observed by the PAMELA collaboration. When the dark matter mass
is 300 GeV (top left panel), 600 GeV (top right panel) and 1000 GeV (bottom
panel), the corresponding lifetimes are 
$\tau_{\rm DM}=4.0,~2.3,~1.6\times10^{26}$, respectively. 
The dashed lines indicate the total fluxes for the MIN propagation model, 
the dotted lines for the MED propagation model, 
and the dash-dotted lines for the MAX propagation model 
({\it cf.} Table \ref{tab:param-antideuteron}). We also show the 
purely secondary fluxes as solid lines.
}
\end{figure}

We show in Fig.~\ref{WePAMELA} the total antiproton and antideuteron fluxes
from the decay of a fermionic dark matter particle in the channel 
$\psi \rightarrow W^\pm e^\mp$, fixing for each dark matter mass the
lifetime in order to account for the steep rise in the positron
fraction observed by PAMELA. For a dark matter mass 
$m_{\rm DM}=300,~600,~1000$ GeV, the corresponding lifetimes are 
$\tau_{\rm DM}=~4.0,~2.3,~1.6\times 10^{26}$ s, respectively.
For these particularly interesting decaying dark matter
scenarios, we find that the antideuteron flux could be within the reach of
the planned experiments AMS-02 and GAPS, provided the dark matter particle
is not too heavy. The discovery of antideuterons would thus favor
the decay mode $\psi\rightarrow W^\pm e^\mp$ as a possible origin of the
PAMELA positron excess over the purely leptonic decay modes such as
$\psi\rightarrow  e^+ e^- \nu$.\footnote{The decay mode 
$\psi\rightarrow W^\pm e^\mp$ may also yield signatures in the 
diffuse gamma-ray background.}
It is important to emphasize that this conclusion holds for ranges
of propagation parameters which not only reproduce the positron
excess observed by PAMELA, but also yield a total antiproton flux
consistent with present measurements.

\section{Conclusions}

We have calculated the antideuteron fluxes at Earth from the decay
of dark matter particles in the Milky Way halo and we have discussed
the prospects to observe cosmic antideuterons in the projected
experiments AMS-02 and GAPS. We have adopted a model-independent
approach analyzing possible hadronic decay modes for the dark matter 
particle. The nuclear fusion of an antiproton and an antineutron
to form antideuteron was simulated employing the coalescence model,
while the propagation of antideuterons in the Galaxy was described by
a stationary two-zone diffusion model with cylindrical boundary 
conditions. The hadronic showers also produce a flux of primary
antiprotons, which is severely constrained by observations. Therefore,
we have simultaneously calculated the predicted antiproton flux, 
to verify whether the chosen parameters and decay modes are consistent 
with present observations.

We have shown that there are choices of parameters where the
antideuteron flux from dark matter decay can be much larger
than the purely secondary component from spallation of cosmic
rays on the interstellar medium, while at the same time the
total antiproton flux remains consistent with observations. We have also
shown that if the dark matter particle is sufficiently light,
the antideuteron flux from dark matter decay could be within the reach
of the planned experiments AMS-02 or GAPS, while the secondary
component is expected to lie below the sensitivity of planned
experiments. Therefore, the observation of cosmic antideuterons in
the near future could be interpreted as an indication for hadronic
decays of dark matter particles. In particular, this conclusion holds 
for a fermionic dark matter particle which decays preferentially via
$\psi\rightarrow W^\pm e^\mp$, which has been proposed as an explanation
for the steep rise in the positron fraction observed by the PAMELA
collaboration.

\section*{Acknowledgements}
This work was partially supported by the DFG cluster of excellence 
``Origin and Structure of the Universe.''

\end{document}